\documentstyle[12pt,psfig]{article}

\def\be{\begin{equation}}
\def\ee{\end{equation}}
\def\bea{\begin{eqnarray}}
\def\eea{\end{eqnarray}}

\newcommand{\beq}{\vspace{2mm}\begin{eqnarray}}
\newcommand{\eeq}{\end{eqnarray}\vspace{2mm}}

\newcommand{\nn}{\nonumber}

\newcommand{\ovm}{{\overline m}}

\def\be{\begin{equation}}
\def\ee{\end{equation}}
\def\bea{\begin{eqnarray}}
\def\eea{\end{eqnarray}}
\begin{document}

\begin{titlepage}
{\sf
\begin{flushright}
{TUM--HEP--265/96}\\
{SFB--375/139}\\
{December 1996}
\end{flushright}
}
\vfill
\begin{center}
{\large \bf {Bottom--up Approach in Supersymmetric
Models$^{\mbox{\boldmath $\ast$}}$}}

\vskip1.5cm
{\sc Marek Olechowski${}^\dagger$}\\
\vskip 1.2cm
{\em Institut f\"{u}r Theoretische Physik} \\
{\em Physik Department T30} \\
{\em Technische Universit\"at M\"unchen} \\
{\em D--85747 Garching} \\
{\em Germany}
\end{center}
\vfill

\thispagestyle{empty}

\begin{abstract}
We present a bottom-up approach to the question of supersymmetry breaking in
the MSSM. Starting with the experimentally measurable low energy supersymmetry
breaking parameters which can take any values consistent with present
experimental constraints, we evolve them up to an arbitrary high energy scale.
Approximate analytical expressions for such an evolution, valid for low and
moderate values of $\tan\beta$, are presented.
We then discuss qualitative properties of the high energy parameter space
and in particular, identify the conditions on the low energy spectrum which
are necessary for the parameters at high energy scale to satisfy simple
regular pattern such as universality or partial universality.
\end{abstract}

\hrule width 5.cm \vskip 1.mm
{\small\small  $^\ast$ Talk given at the 28th International Conference 
on High Energy Physics, Warsaw, Poland, July 1996}

{\small\small ${}^\dagger$ On leave of absence from the
Institute of Theoretical Physics, Warsaw University,  Warsaw, Poland}
\end{titlepage}

\section{Introduction}

In supersymmetric theories the mechanism of supersymmetry breaking
remains a fundamental open problem. Its low energy manifestation is
the supersymmetric spectrum. Therefore, one may hope to get some
insight into this problem from experiment. Currently, the most popular
view on supersymmetry breaking is that the parameters
of the low energy effective theory have their origin in the GUT (or
string) scale physics \cite{NILLES}. 
Supersymmetry is spontaneously broken in an
invisible sector and this effect is transferred to our sector through
supergravity interactions at the scale $M_{Pl}$.
Other models  have also been proposed \cite{DINE} in which the supersymmetry
breaking is communicated to the electroweak sector through some other
messengers at the energy scale $M\ll M_{Pl}$.
In the Minimal Supersymmetric Standard Model (MSSM) the soft supersymmetry
breaking parameters at the large scale
are connected to their low energy values via the renormalization group
equations (RGE) which do not contain any new unknown parameters.
With any simple theoretical {\sl Ansatz} for the pattern of soft supersymmetry
breaking terms at the scale where supersymmetry breaking is transmitted
to the observable sector one can study superpartner spectra in the top-down
approach. It is clear that any particular {\sl Ansatz} for parameters
at high energy scale gives only a small subspace 
of all low energy parameter space. In order to have a broad overview of the
low energy - high energy parameter mapping
it is of interest to supplement the top-down approach
with a bottom-up one where we can learn how certain qualitative features
of the low energy spectra reflect themselves on the
qualitative pattern of soft terms at different energy scales. Of course,
eventual measurement of the superpartner and Higgs boson spectra
and various mixing angles in the sfermion sector will permit (in the
framework of MSSM)  a complete bottom$-$up mapping. We shall know
then the pattern of the soft parameters at any hypothetical scale $M$
of supersymmetry breaking and this will have major impact on our ideas
on its origin.

This talk is based on results obtained in reference
\cite{CCOPW}.
We present here the main features of
the bottom$-$up mapping for the set of
parameters $\mu$, $M_2$, $m^2_{H_1}$, $m^2_{H_2}$,
$B$ and the third generation squark mass
parameters $m^2_Q$, $m^2_U$, $m^2_D$ and $A_t$.
To a very good approximation this is a closed set of
parameters, whose RG running decouples from the remaining
parameters
\footnote{In the small or moderate $\tan\beta$ regime,
           the dependence of the RG running of these parameters
           on slepton (and the two first generation sfermion) masses
           comes only through the  small hypercharge D-term contributions
           which we neglect here. Their contribution can be found in
           \cite{CCOPW}.}.
We shall concentrate on the region of small to moderate values
of $\tan\beta$ in  which, for $m_t = 175 \pm6$ GeV, the bottom quark
Yukawa coupling effects may be neglected.
For this case we present analytic expressions (at one loop level) for
the values of our set at $M_{GUT}$ and at any $M < M_{GUT}$ in terms of
its values at $M_Z$ scale. The equations are valid for arbitrary boundary
values of the parameters at the scale $M$.

In absence of direct experimental measurement of the low
energy parameters, we disscuss the mapping of the region
characterized by light chargino and right handed
stop with masses of the order of $M_Z$ which
is consistent with all existing
experimental constraints and is of interest for LEP2.

The bottom--up mapping is presented in section 2.
Section 3 containes
some general results obtained with this method
and its application to the low energy region
with light chargino and stop.
More detailed presentation of the bottom--up mapping
and more results
can be found in reference \cite{CCOPW}.

\section{Solutions to the renormalization group equations}

We write down first the approximate solutions to the RG equations in
the low and moderate $\tan\beta$ regime
(i.e. neglecting the effects of Yukawa couplings other
than the top quark one) for the relevant
parameters. For our purpose, it is useful to give the expression of the
high energy parameters as a function of the low energy ones. They read:
\begin{eqnarray}
\ovm^2(0) 
&=& {\ovm^2(t)\over 1-y} + y\left(A_t^2(0)-2\hat\xi A_t(0)M_{1/2}\right)
+ {y\hat\eta - y^2\hat\xi^2 - \overline\eta\over 1-y} M_{1/2}^2
\label{eqn:m2ov0}\\
m_K^2(0) &=& m_K^2(t) + {c_K\over c}{y\over 1-y}\ovm^2(t)
+\frac{c_K}{c} y \left(A_t^2(0) - 2 \hat\xi A_t(0)M_{1/2}\right)
\nn\\
&-& \left[\eta_K + \frac{c_K}{c}{y\over 1-y}\left(\overline\eta
- \hat\eta  + y \hat\xi^2 \right)\right]M_{1/2}^2
\label{eqn:m2k0}\\
A_t(0) &=& \frac{A_t(t) + \left(\xi_u - y\hat\xi\right) M_{1/2}}{1-y}
\label{eqn:at0}
\\
B(0)
&=& B(t) + \frac{c_B}{c}\frac{y}{1-y}A_t(t)
+\left[\xi_B +
\frac{c_B}{c}\frac{y}{1-y}\left(\xi_u-{\hat\xi}\right)\right] M_{1/2}
\label{eqn:b0}
\\
\mu^2(0)&=&
{\mu^2(t) \over \left(1-y\right)^{\left({{c_\mu}/{c}}\right)}}
\prod_{i=1,2,3}
\left({\alpha_i(t)\over\alpha_i(0)}
\right)^{\left({a^i_{\mu} / b^i}\right)}
\label{eqn:mu0}
\end{eqnarray}
where $\ovm^2(t)\equiv m^2_Q(t) + m^2_U(t) + m^2_{H_2}(t)$,
$\overline\eta\equiv\eta_Q+\eta_U+\eta_{H_2}$,
$t\equiv{1\over2\pi}\log{M\over Q}$, $M=M_{GUT}$ or
any intermediate scale, and $m_K^2$, with $K = H_i,Q,U,D$, denotes
the Higgs, left handed squark, right handed up-type squark
and right handed down-type squark
soft supersymmetry breaking mass parameters, respectively.
Quantities at $t=0$ are the initial
values of the parameters at the scale $M$. The
coefficients $c$ and $c_K$ read:
$c=6$, $c_B=3$, $c_Q=1$,  $c_U=2$, $c_{H_2}=3$,
$c_D=c_{H_1}=0$; $c_{\mu}=3$.
The function ~$y\equiv y(t)$ ~is defined as
\begin{eqnarray}
y(t) \equiv {Y_t(t)\over Y_f(t)}.
\label{eqn:y}
\end{eqnarray}
where $Y_t = h_t^2/4\pi$ ($h_t$ is the top quark Yukawa coupling)
have the well known form:
\begin{equation}
Y_t(t)= \frac{Y_t(0)E(t)}{1+c Y_t(0) F(t)}
\label{eqn:yt}
\end{equation}
and its infrared fixed point value, $Y_f(t)$,
is given by
\begin{equation}
Y_f(t)= {{E(t)}\over{c F(t)}}
\label{eqn:yf}
\end{equation}
with
\begin{eqnarray}
E(t) &=& \prod_{i=1,2,3}
\left(\frac{\alpha_i(0)}{\alpha_i(t)}\right)^{\left(a^i/b^i\right)}
\\
F(t) &=& \int_0^t E(t^{\prime})dt^{\prime}.
\label{eqn:fft}
\end{eqnarray}
Functions $\xi_j(t)$, $\eta_K(t)$, $\hat\xi(t)$, $\hat\eta(t)$
are defined by
\begin{eqnarray}
\xi_j(t) &=&
I\left[\sum_i \frac{M_i(0)}{M_{1/2}}\frac{a_j^i\alpha_i^2(t)}{\alpha_i(0)}
\right]
\\
\eta_K(t) &=&
I\left[\sum_i\frac{M_i^2(0)}{M_{1/2}^2}\frac{d_K^i\alpha_i^3(t)}{\alpha_i^2(0)}
\right]
\\
{\hat\xi}(t) &=&
H\left[\sum_i \frac{M_i(0)}{M_{1/2}}\frac{a^i_U\alpha_i^2(t)}{\alpha_i(0)}
\right]
\\
{\hat\eta}(t)
&=&
H\left[\sum_i\frac{M_i^2(0)}{M_{1/2}^2}
\frac{\overline d^i\alpha_i^3(t)}{\alpha_i^2(0)}\right]
\nn\\
&+&
H\left[\sum_i\frac{M_i(0)}{M_{1/2}}
\frac{\overline d^i\alpha_i^2(t)}{\alpha_i(0)}\xi_u(t)\right]
\end{eqnarray}
where the coefficients $a^i_j$ and $d^i_j$
read: $a_u^i = (13/15, 3, 16/3)$,
$a_d^i = (7/15, 3, 16/3)$, $a_B^i = (3/5, 3, 0)$;
$a^i_{\mu}=(5/3,3,0)$;
$d_Q^i = (1/15, 3, 16/3)$, $d_U^i = (16/15, 0, 16/3)$,
$d_D^i = (4/15, 0, 16/3)$, $d^i_{H_1} = d^i_{H_2} = (3/5, 3, 0)$,
$\overline d^i\equiv d^i_Q + d^i_U + d^i_{H_2}$ while $H$ and $I$
are defined in the following way:
\begin{eqnarray}
H[f(t)] &=& \int_0^t f(t^{\prime}) dt^{\prime}
- \frac{1}{F(t)}\int_0^t F(t^{\prime})f(t^{\prime})dt^{\prime}
\nn\\
I[f(t)] &=& \int_0^t f(t^{\prime})dt^{\prime}
\end{eqnarray}
Factors $M_i(0)/M_{1/2}\neq1$ appear because we do not assume exact
gauge coupling unification and $M_{1/2}$ can be chosen
by convention e.g.\ as $M_{1/2}\equiv M_3(0)$.
Parameters $\xi_j$ and $\eta_K$ can be computed analytically
but $\hat\xi$ and $\hat\eta$ require numerical integration.
We give their typical values in table 1. More values of
all relevant parameters can be found in reference
\cite{CCOPW}.
\begin{table}\begin{center}
\vspace{0.4cm}
\begin{tabular}{| c | c | c | c | c | c |} \hline
\multicolumn{2}{|l|}{~$\hat\xi$}&\multicolumn{4}{c|}{$\,M$ [GeV]}\\
\cline{3-6}
\multicolumn{2}{|l|}{}&
    $2\cdot10^{16}$ & $1\cdot10^{10}$ & $1\cdot10^7$    & $1\cdot10^5$\\
\hline
&   .115   &     2.16      &   1.07      &    .640       &   .375        \\
\cline{2-6}
$\alpha_3$
&   .120   &     2.23      &   1.11      &    .664       &   .389        \\
\cline{2-6}
&   .125   &     2.30      &   1.15      &    .687       &   .403        \\
\hline
\end{tabular}

\begin{tabular}{| c | c | c | c | c | c |}
\hline
\multicolumn{2}{|l|}{~$\hat\eta$}&\multicolumn{4}{c|}{$\,M$ [GeV]}\\
\cline{3-6}
\multicolumn{2}{|l|}{}&
    $2\cdot10^{16}$ & $1\cdot10^{10}$ & $1\cdot10^7$    & $1\cdot10^5$\\
\hline
&   .115   &     12.2      &   4.16      &    2.00       &   .988        \\
\cline{2-6}
$\alpha_3$
&   .120   &     12.8      &   4.40      &    2.11       &   1.04        \\
\cline{2-6}
&   .125   &     13.5      &   4.64      &    2.22       &   1.09        \\
\hline
\end{tabular}
\end{center}

\begin{center}
{\sl Table 1: Typical values of $\hat\xi$ and $\hat\eta$.}
\end{center}
\end{table}

The soft SUSY breaking
parameters are expressed in terms of the physical parameters according to
\begin{eqnarray}
m^2_{H_1}(t) &=& \sin^2\beta M^2_A + {t_{\beta}\over2} M^2_Z - \mu^2
\\
m^2_{H_2}(t) &=& \cos^2\beta M^2_A - {t_{\beta}\over2} M^2_Z - \mu^2
\\
\mu(t) B(t) &=& \sin\beta\cos\beta M^2_A
\\
m^2_Q(t) &=& M^2_{\tilde t_1}\cos^2\theta_{\tilde t}
         + M^2_{\tilde t_2}\sin^2\theta_{\tilde t}
         - m^2_t
         - {t_{\beta}\over6}\left(M^2_Z - 4M^2_W\right)
\\
m^2_U(t) &=& M^2_{\tilde t_1}\sin^2\theta_{\tilde t}
         + M^2_{\tilde t_2}\cos^2\theta_{\tilde t}
         - m^2_t
         +{2\over3}t_{\beta}\left(M^2_Z - M^2_W\right)
\\
A_t(t) &=& {M^2_{\tilde t_1}-M^2_{\tilde t_2}\over m_t}\sin\theta_{\tilde t}
\cos\theta_{\tilde t}
       + \mu(t)\cot\beta
\label{eqn:att}
\end{eqnarray}
where $M^2_{\tilde t_1}$, $M^2_{\tilde t_2}$ and $\theta_{\tilde t}$
are respectively the heavier and lighter physical top squark masses and their
mixing angle while $t_{\beta}\equiv(\tan^2\beta-1)/(\tan^2\beta+1)$.
%
%We have ignored the low energy one-loop corrections to the mass
%parameters, which are not essential to determine the qualitative
%properties of the mass parameters at the scale $M$.
%
One should stress that eqs.\ (\ref{eqn:m2ov0}-\ref{eqn:mu0})
are valid for general, non-universal values of the soft SUSY breaking mass 
parameters at the scale $M$ and that unification assumptions for the
gauge couplings and for gaugino masses have not been used
($M_{1/2}$ is by convention equal to the gluino mass $M_3$ at the
scale $M$). Moreover, the functions $y(t)$ and $Y_f(t)$ defined by eqs.\
(\ref{eqn:y},\ref{eqn:yf}) are auxiliary functions defined
for any scale $M$.
The function $y(t)$ is very convenient in presenting the solutions
to the RG equations. The whole dependence of the results on the
large top Yukawa coupling can be easily expressed in terms of $y$
(see eqs.\ (\ref{eqn:m2ov0}--\ref{eqn:mu0})). For $M=M_{GUT}$ a consistent
perturbative treatment of the theory can only be performed if
\begin{equation}
y_{GUT}\equiv y\left(t={1\over2\pi}\log{M_{GUT}\over M_Z}\right) < 1,
\end{equation}
where $y_{GUT}\approx1$ defines the quasi infrared fixed point solution.
%
%In this limit we obtain the well known
%dependence of $m_t$ on $\tan\beta$ which is consistent with
%perturbativity up to the GUT scale.
%
For scenarios with supersymmetry broken
at lower scales $M \ll M_{GUT}$, the same
values of $m_t$ and $\tan\beta$ (or equivalently of
$y_{GUT}$) give obviously much lower values for the auxiliary
function $y(t)$, where $t$ is defined at the scale $Q=M_Z$,
$(t={1\over2\pi}\log{M\over M_Z})$.
One should also remember that, in general,
for $M < M_{GUT}$, new matter multiplets are expected to
contribute to the running of the gauge
couplings above the scale $M$. New complete $SU(5)$ or $SO(10)$
multiplets do not destroy the unification of gauge couplings
but their value at the unification scale becomes larger and,
correspondingly, the perturbativity bound, $y_{GUT}<1$, allows for
larger values of the top quark Yukawa coupling (smaller $\tan\beta$)
at the $M_Z$ scale. We shall  use the auxiliary function $y(t)$
when presenting our results.
\vskip 0.5cm

\section{Results of bottom--up mapping in MSSM.}

We recall that our theoretical intuition about the superpartner spectrum
has been to a large extent developed on the top-down approach in
the minimal supergravity model (with universal soft terms) and on
minimal models of low energy supersymmetry breaking.
For instance, it is well known that in the minimal supergravity model
for low $\tan\beta$ the renormalization group running gives
\footnote{The coefficients in eq.\ (\ref{eqn:mz_univ})
               are for $\tan\beta=1.6$; they decrease with
               increasing $\tan\beta$.
               The exact values of the coefficients in eqs.\
               (\ref{eqn:mz_univ},\ref{eqn:mq_univ})
               depend slightly on the values of $\alpha_s$,
               $\sin^2\theta_W$ and $M_{GUT}$ chosen.}:
\begin{eqnarray}
M^2_Z\approx-2\mu^2 + {\cal O}(3) m^2_0 + {\cal O}(12) M^2_{1/2}
\label{eqn:mz_univ}
\end{eqnarray}
where $M_{1/2}$ and $m_0$ are the universal gaugino and scalar mass
parameters (at the high energy scale) respectively. The large 
coefficient in front of $M^2_{1/2}$ is the well known source of
fine tuning for $M_{1/2}>M_Z$ and therefore one naturally expects
light neutralino and  chargino both gaugino-like since, from 
(\ref{eqn:mz_univ}), $\mu>M_{1/2}$ \cite{COPW}.
For moderate values of the parameters at the GUT scale,
the RG evolution of the squark masses gives
\begin{eqnarray}
m^2_Q&=&{\cal O}(6) M^2_{1/2} + {\cal O}(0.5) m^2_0 + \ldots
\nonumber\\
m^2_U&=&{\cal O}(4) M^2_{1/2} + \ldots
\label{eqn:mq_univ}
\end{eqnarray}
where ellipsis stand for terms proportional to $A_0^2$ and $A_0 M_{1/2}$ 
with coefficients going to zero in the limit $y \to 1$. 
We see that the hierarchy $m^2_Q \gg m^2_U$ can be generated provided
${\cal O}(1 {\rm TeV}) \sim m_0 \gg M_{1/2} \sim {\cal O}(M_Z)$, i.e.
consistently with naturality criterion.
For large enough values of
the universal trilinear soft terms, $A_0$, one
may even obtain $m^2_U<0$ and in consequence $M_{\tilde t_1} \gg
M_{\tilde t_2} \sim {\cal O}(M_Z)$, $m_{C_1^+}\sim {\cal O}(M_Z)$ and
gaugino-like \cite{STRUMIA}.

Of course,  {\sl Ans\"atze} for soft terms at high 
energy scale such as full
universality or partial universality selects only a small subset of the
low energy parameter space, even if the latter is assumed to contain
light chargino and stop and to be characterized by the hierarchy
$M_{\tilde t_1} \gg M_{\tilde t_2}$. Since the sparticle spectrum with
$m_{\chi^+_1}\sim m_{N_1^0} \sim M_{\tilde t_2} \sim {\cal O}(M_Z)$
and $M_{\tilde t_1} \gg M_{\tilde t_2}$ is important for LEP2 and
at the same time consistent with the precision electroweak data
\cite{CP},
it is of interest to study the mapping of such
a spectrum to high energies in a general way and to understand better its
consistency (or inconsistency) with various simple patterns.

We shall discuss now some general features of the behaviour of the soft
supersymmetry breaking parameters, which may be extracted from
eqs.\ (\ref{eqn:m2ov0}-\ref{eqn:mu0}).
We consider first the case $M=M_{GUT}$ and
the limit $y_{GUT}\rightarrow1$ as a useful reference frame. One
should be aware, however, that already for $y_{GUT}\simeq 0.8$--$0.9$ large
qualitative departures from the results associated with
this limit are possible.
In the limit $y_{GUT}\rightarrow1$, from eqs.\ (\ref{eqn:m2ov0}) and
(\ref{eqn:at0}), we see that if the high energy parameters are of the same
order of magnitude as the low energy ones, the following relations must
be fulfilled:
\begin{eqnarray}
\Delta_{m^2}&\equiv&
\ovm^2(t)-\left(\overline\eta -y\hat\eta
+ y^2 \hat{\xi}^2\right)M^2_{1/2}\rightarrow0\nn\\
\Delta_A&\equiv&A_t(t)-\left(y\hat\xi -\xi_u\right) M_{1/2}\rightarrow0 ,
\label{eqn:sumrule}
\end{eqnarray}
irrespectively of their initial values (IR fixed point). This is a well 
known prediction, which, for $y_{GUT}\neq1$, remains  valid in  the
gaugino dominated supersymmetry breaking scenario in the minimal 
supergravity model.

Let us now suppose that the relations (\ref{eqn:sumrule}) are strongly violated
by experiment, namely the scalar masses (or the soft supersymmetry
breaking parameter $A_t$) are much different than predicted by
(\ref{eqn:sumrule}), for the values of $m_t$ and $\tan\beta$ corresponding
to $y_{GUT}$ close to 1. Clearly, this means that
\begin{eqnarray}
A_t(0)&\sim & {\cal O}\left({\Delta_A\over1-y}\right)\nn \\
\overline m^2(0)
&\sim &{\cal O}\left({\Delta_{m^2}\over1-y}\right)+ {\cal O}(A_t^2(0))  
\end{eqnarray}
i.e. supersymmetry breaking must be driven by very large initial scalar masses
and the magnitude of the effect depends on the departure from the 
fixed point relations for $A_t$ and $\ovm^2$, eq.\
(\ref{eqn:sumrule}), and the proximity to the top quark mass infrared
fixed point solution measured by $1-y_{GUT}$. Moreover, for not very
small values of $|\Delta_A|$ and/or $|\Delta_{m^2}|$ it follows
from eq.\ (\ref{eqn:m2k0}) that in the limit $y_{GUT}\rightarrow1$,
independently of the actual
values of the masses at the scale $M_Z$,
their initial values are correlated such that
\begin{equation}
m^2_Q(0) : m^2_U(0) : m^2_{H_2}(0) \simeq 1 : 2 : 3
\label{eqn:fprel}
\end{equation}
and $m^2_{H_1}(0)$ is much smaller than the above three soft masses.
In other words, the values at
$M_Z$ are obtained (via RGEs) by a very high degree
of fine tuning between the initial values $m^2_K(0)$.
We conclude that eqs.\ (\ref{eqn:sumrule}) are necessary (but of course
not sufficient) conditions for large departures from the prediction
(\ref{eqn:fprel}) for the soft scalar masses in the limit $y\rightarrow1$.
In particular, they are necessary conditions for the spectrum consistent
with fully (or partially) universal initial values of the third
generation squark and Higgs boson soft masses. 

When eqs.\ (\ref{eqn:sumrule}) are violated,
there are essentially two ways of departing from the prediction
(\ref{eqn:fprel}). One is to increase the value of $\tan\beta$
(i.e. to decrease the top quark Yukawa coupling for fixed $m_t$).
The other way is to lower the scale at which supersymmetry breaking is 
transferred to the observable sector, since for $M\ll M_{GUT}$ the
soft supersymmetry breaking parameters do not feel the strong rise
of the top quark Yukawa coupling at scales close to $M_{GUT}$. In both
cases, $y$ takes values smaller than one and we can depart from
(\ref{eqn:fprel}) even when (\ref{eqn:sumrule}) are not satisfied.

We shall illustrate the above general considerations by numerical 
bottom--up mapping of the low energy parameter space characterized 
as follows: $m_{\chi^+} = 90$ GeV, $M_{\tilde t_2} = 60$ GeV
(the pattern of soft supersymmetry breaking parameters does not
depend strongly on the exact value of the light chargino and stop masses),
$0.1 < M_2/|\mu| < 10$ for both signs of $\mu$.
To reduce the parameter space, in our numerical analysis we make the
assumption that
${M_1/\alpha_1}={M_2/\alpha_2}={M_3/\alpha_3}$
at any scale. For $m_t=175$ GeV, $\alpha_s(M_Z)=0.118$ 
we consider two examples:
$M=M_{GUT}=2\times10^{16}$ GeV,  
$y_{GUT} = 0.98$ (corresponding to $\tan\beta \approx 1.6$)
and 
$M=10^{7}$ GeV,
$y = 0.75$ (corresponding to $\tan\beta \approx 1.25$).

We scan over the remaining relevant parameters $M_A$,
$M_{\tilde t_1} < 1$ TeV, and the whole range of $\theta_{\tilde t}$. 
The accepted low energy parameter space is then defined as a subspace
in which $M_h>60$ GeV, $0.98\times10^{-4} < BR(b\rightarrow s\gamma) 
<3.66\times10^{-4}$, and we also 
require that $\chi^2\leq\chi^2_{min ~SM} + 2$
where $\chi^2$ is for the fit to the elctroweak observables which do not
involve heavy flavors.

\begin{figure*}
\psfig{figure=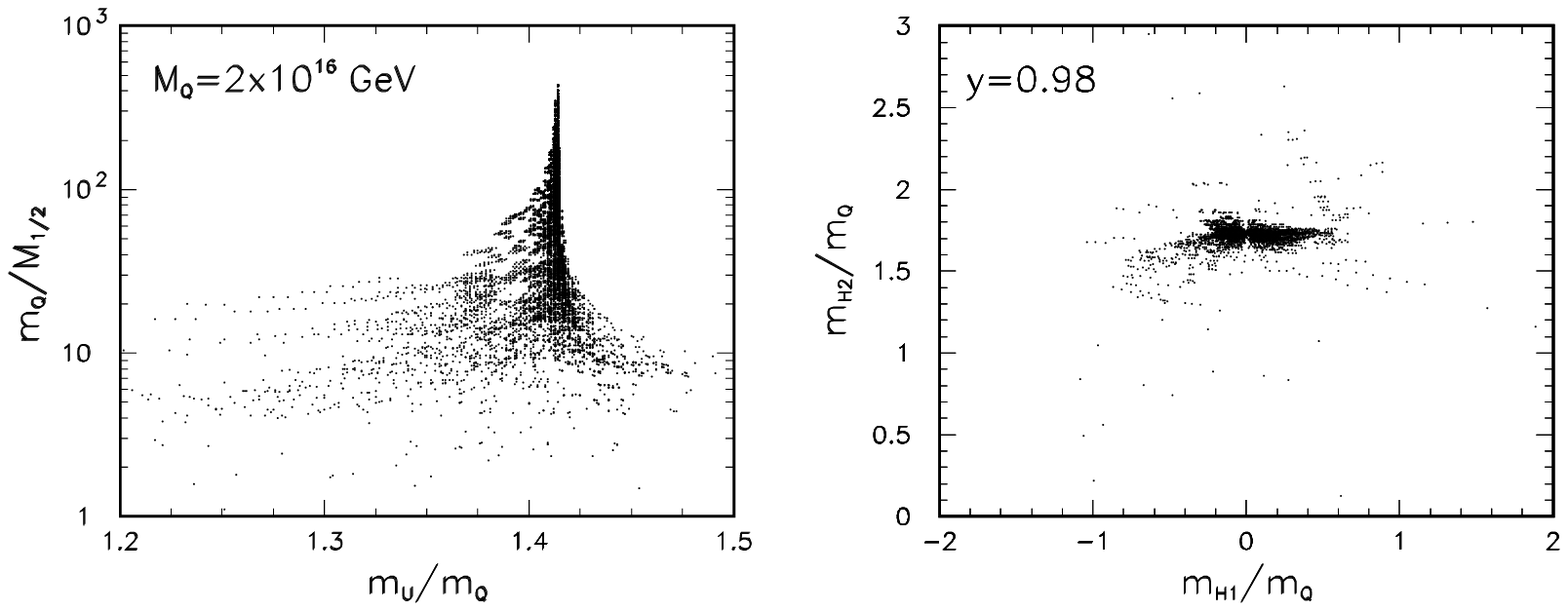,height=2.3in}
\vskip3mm
{\sl Figure 1: Soft supersymmetry breaking parameters at the scale
$M=M_{GUT}=2\cdot 10^{16}$ GeV, for $y=0.98$ obtained by mapping the low energy
parameter space specified in the text.}
\vskip2mm
\end{figure*}

In Figure 1 we show the dependence of $m_U/m_Q$
on the value of $M_{1/2}$ and
the behaviour of the soft supersymmetry breaking
Higgs mass parameters in the infrared limit: $y = 0.98$
(masses are defined by $m_K=m_K^2/\sqrt{|m_K^2|}$).
A clear concentration of solutions around $m_U/m_Q = \sqrt{2}$
and $m_{H_2}/m_Q = \sqrt{3}$ appears. The above ratios are 
governed by the size of the gaugino masses.
For low values of $M_{1/2}$, the maximum value
of $m_U/m_Q$ is given by $\sqrt{2}$, while for solutions with large
values of $M_{1/2} = {\cal O}$(1 TeV), $m_U/m_Q > \sqrt2$ is possible.
Similarly, values of $m_{H_2}/m_Q\geq\sqrt3$ may only appear for large
values of the gaugino masses, while low values of the gaugino masses
always lead to $m_{H_2}/m_Q\leq\sqrt{}3$. For $y\simeq 1$, 
a strong concentration of solutions around the boundary values is
observed. Such behaviour of the solutions is just
a reflaction of the properties discussed above for $y$ close to 1
and shows the global tendency in the mapping of the low energy region
selected by our criteria.

\begin{figure*}
\psfig{figure=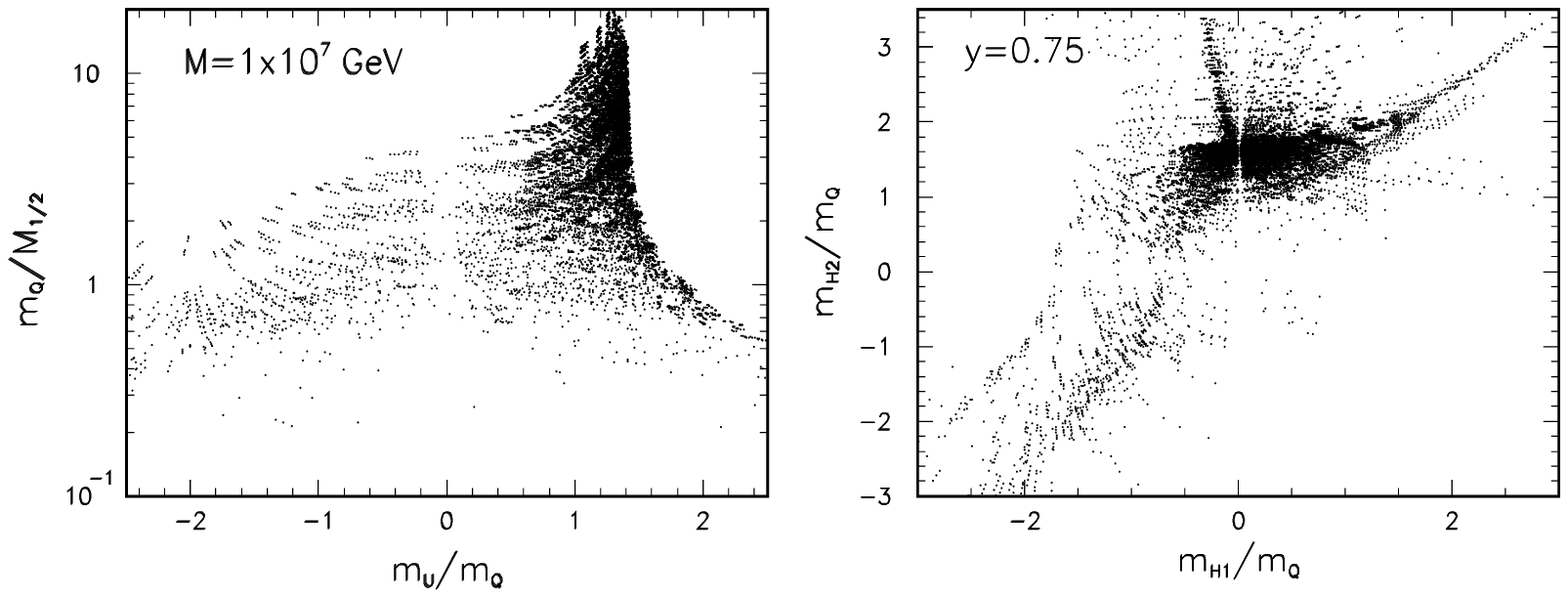,height=2.3in}
\vskip3mm
\begin{center}
{\sl Figure 2: Same as Figure 1, but for $M=10^7$ GeV, $y$=0.75.}
\end{center}
\vskip2mm
\end{figure*}

Figure 2 shows the solutions obtained for low supersymmetry
breaking scale $M=10^7$ GeV and for $y=0.75$. The value of $y$
is now much smaller than 1 and the violation of the sum rules,
eq.\ (\ref{eqn:sumrule}),
does not imply an  enhancement of the values of the parameters
at the scale $M$. From the analysis of our equations, for these low
values of $M$ and $y(t)$, it follows that
the initial values of soft terms at the
scale $M$
tend to reflect the pattern observed at the $M_Z$ scale.
Due to the small renormalization group running effects
and the fact that $\eta_Q\simeq 0.85$, the ratio
$m_Q/M_{1/2}$ is just a reflection of the chosen values of
the left handed stop parameter and the gluino mass at
low energies, increasing for lower values of $M_{1/2}$.
More interesting is the behaviour of $m_U/m_Q$.
The concentration of solutions around
$m_U/m_Q \simeq \sqrt{2}$ disappears for this low values of
$y$ and, instead, the ratio $m_U/m_Q$ tends, in general, to be lower.
Hence, a light stop is not necessarily
in conflict with models of supersymmetry breaking in which
$m_U/m_Q\simeq 1$ at the scale $M$.
Figure 2 shows that for SUSY breaking scale of order $10^7$ GeV,
the three soft masses $m_Q$, $m_U$ and $m_{H_2}$ can have
similar values for many solutions, especially for light gauginos.
The second soft Higgs mass, $m_{H_1}$, tends to be rather light
but the full universality can also be easily obtained.

\section{Conclusions}

In this paper we have discussed the mapping to high energies of the low
energy parameters of the MSSM.
We investigated the specific region in low energy parameter space
with light chargino and right-handed stop which can be interested
for LEP2.
For heavy top quark and small or moderate $\tan\beta$
the global pattern of the mapping
is determined by the proximity of the top quark Yukawa coupling to its
IR fixed point value and by the assumed scale at which supersymmetry
breaking is transmitted to the observable sector.
The general pattern of this mapping is the dominance
of the scalar masses (over the gaugino mass) in the supersymmetry
breaking. We have also identified
the conditions which are necessary for the spectrum to be consistent
with a  simple {\sl Ansatz} like universality (or partial universality) of the
high energy values for the scalar masses. In particular, the $SO(10)-$type
initial conditions, with universal squark masses
but with non-universal Higgs masses are compatible with an interesting
subregion of the considered parameter space.

\section*{Acknowledgements}

This work was partly supported by Deutsche Forschungsgemeinschaft
grant SFB--375/95,
by the European Commission TMR programmes ERBFMRX-CT96-0045
and ERBFMRX-CT96-0090
and by Polish Committee for Scientific Research.

\end{document}